\documentclass[11pt,twoside]{article}
\usepackage{asp2014}

\aspSuppressVolSlug
\resetcounters

\begin{document}

\title{INAF Trieste Astronomical Observatory\\Information Technology Framework}

\author{S.~Bertocco,$^1$ D.~Goz,$^1$ L.~Tornatore,$^1$ A.~Ragagnin,$^1$ G.~Maggio,$^1$ F.~Gasparo,$^1$ C.~Vuerli,$^1$ G.~Taffoni,$^1$ and M. Molinaro$^1$}
\affil{$^1$Italian National Institute of AstroPhysics, Osservatorio Astronomico di Trieste, Italy; \email{sara.bertocco@inaf.it}}

\paperauthor{Sara~Bertocco}{sara.bertocco@inaf.it}{0000-0003-2386-623X}{inaf}{OATs}{Trieste}{}{34143}{Italy}
\paperauthor{David~Goz}{david.goz@inaf.it}{0000-0001-9808-2283}{INAF}{OATs}{Trieste}{}{34143}{Italy}
\paperauthor{Luca~Tornatore}{luca.tornatore@inaf.it}{0000-0003-1751-0130}{INAF}{OATs}{Trieste}{}{34143}{Italy}
\paperauthor{Antonio~Ragagnin}{antonio.ragagnin@inaf.it}{0000-0002-8106-2742}{INAF}{OATs}{Trieste}{}{34143}{Italy}
\paperauthor{Gianmarco~Maggio}{gianmarco.maggio@inaf.it}{0000-0003-4020-4836}{INAF}{OATs}{Trieste}{}{34143}{Italy}
\paperauthor{Federico~Gasparo}{federico.gasparo@inaf.it}{0000-0002-1591-5428}{INAF}{OATs}{Trieste}{}{34143}{Italy}
\paperauthor{Claudio~Vuerli}{claudio.vuerli@inaf.it}{0000-0002-9640-8785}{INAF}{OATs}{Trieste}{}{34143}{Italy}
\paperauthor{Giuliano~Taffoni}{giuliano.taffoni@inaf.it}{0000-0001-7011-4589}{INAF}{OATs}{Trieste}{}{34143}{Italy}
\paperauthor{Marco~Molinaro}{marco.molinaro@inaf.it}{0000-0001-5028-6041}{INAF}{OATs}{Trieste}{}{34143}{Italy}




\begin{abstract}
INAF Trieste Astronomical Observatory (OATs) has a long tradition in information 
technology applied to Astronomical and Astrophysical use cases, particularly for 
what regards computing for data reduction, analysis and simulations; data and 
archives management; space missions data processing; design and software development 
for ground-based instruments.
The ensemble of these activities, in the last years, pushed the need to acquire
new computing resources and technologies and to deep competences in theirs management.
In this paper we describe INAF\text{-}OATs computing centre technological stuff,
our involvement in different EU Projects both in the path of building 
of EOSC, the European Open Science Cloud; in the design and prototyping
of new Exascale supercomputers in Europe and the main research
activities carried on using our computing centre.
\end{abstract}
\section{Introduction}
INAF\text{-}OATs commitment in the application of information
technology to Astronomical and Astrophysical use cases
involves a variety of activities.
It participated, since the beginning, in Italian Grid and Cloud initiatives
gaining a major role in Italy and in Europe in large projects for the development of
a multidisciplinary platform for distributed computing and data resources sharing.
Recently it participated in EU EGI.eu projects, coordinating the Astronomy and
Astrophysics community, for the development of a cloud based infrastructure in
Europe with the aim to spread the IVOA standards based interoperability with CANFAR
in Canada.
Now it is active in the design of the SKA regional centres.
INAF\text{-}OATs participats also in different EU projects for the design and prototyping
of new Exascale supercomputers in Europe gaining a leading position. 

To cover the requirements raised by these activities,
INAF\text{-}OATs deployed a computing centre that offers HPC, HTC
and cloud resources for
internal users, INAF and large international projects.
In this paper we describe INAF\text{-}OATs computing centre technological stuff,
our involvement both in the ESCAPE EU Project in the path of building
of EOSC, the European Open Science Cloud, and in the EU projects in the
path toward the exascale era and the main research
activities carried on using our data and computing centre.
\section{Computing and Data Infrastructure}
INAF\text{-}OATs acquired a set of computing resources thanks
to the DHTCS\text{-}IT project (Distributed High Throughput Computing and Storage in Italy)
founded by the Italian "Ministero dell'Istruzione, dell'Universit\`a e della
Ricerca"  and other EU funded projects to
deploy two computing clusters to satisfy different requirements: one targeted
to HPC, named HOTCAT, and another one targeted to HTC, named CloudCAT. 

HOTCAT manages computing resources through PBS 
(Portable Batch System), a job scheduler and workload management software optimized in HPC environments (clusters, clouds, and supercomputers) fast, scalable, secure, and resilient.
It provides storage resources, distributed on 3 nodes, using BeeGFS, a parallel file system 
developed and optimized for high-performance computing, granting a 2 GB/sec of throughput.
It is equipped with an Infiniband interconnect allowing 
high throughput and low latency (1 microsec).
Moreover, it supplys users with more than 60 software environment 
for Astronomical data reduction and analysis and it offers tools for software 
development, profiling and debugging.

CloudCAT provides cloud resources using OpenStack and Openstack Swift as object storage.
Computing and storage resources are provided to users with Virtual  Machine
images pre-configured with Astronomical Software (e.g. ESO Scisoft) and remote 
desktop capabilities to allow easy access and usage.
CloudCAT is compliant with EGI-Federated cloud resources.
\begin{table}[!ht]
\caption{\bf HOTCAT Cluster Hardware}
\smallskip
{\small
\begin{tabular}{ll}  
\noalign{\smallskip}
\tableline
\noalign{\smallskip}
{\bf Computation nodes:} & \\
Cores & 1400 INTEL Haswell E5-4627v3 \\
RAM & 6GB RAM/Core (8.5TB total) \\
Available Storage & 500 TB (BeeGFS) \\
{\bf Storage nodes  :} & \\
Cores & 24 INTEL Haswell E5-4627v3 \\ 
RAM & 256GB RAM per node \\
{\bf Network connection  :} & \\
Infiniband ConnectX -3 Pro Dual QSFP+ 54Gbs & 
\end{tabular}
}
\end{table}
\begin{table}[!ht]
\caption{\bf CloudCAT Cluster Hardware}
\smallskip
{\small
\begin{tabular}{ll}  
\noalign{\smallskip}
\tableline
\noalign{\smallskip}
{\bf Computation nodes   :} & \\
Cores & 200 INTEL Westmere E5620 @ 2.40GHz \\
RAM & 8GB RAM/Core \\
Available Storage & 75 TB Object Storage Swift \\
{\bf Network connection  :} & \\
Infiniband 10Gbs &
\end{tabular}
}
\end{table}
\subsection{The CHIPP project}
The INAF researcher perform tasks like hydrodynamical N\text{-}body simulations; instruments 
performance simulation during designing phases; testing of computing intensive 
analisys algorithms; testing and exploiting software efficiency improvement 
through parallelization of known complex algorithms.
High computing resource intensive programs are generally run 
exploiting specialized infrastructures outside INAF (e.g. GARR) but there
was a lack in resources to fine satisfy small/medium sized programs requirements.
CHIPP born to fill this lack. It is an INAF pilot project to proof the benefit 
to provide Italian Astronomers with a distributed medium sized computing infrastructure,
currently based on already existing resources at Trieste and Catania, and making available
Tier-2/Tier-3  systems (1,200 CPU/core) for all the INAF community. The run programs 
are HPC, data reduction and analysis, machine learning. 
\subsection{Activities towards exascale computing}
INAF is one of the leading institutions participating in different EU projects
aiming at designing and prototyping of new Exascale supercomputers in Europe.

ExaNeSt~\citep{exanest} developed, evaluated, and prototyped the physical 
platform and architectural solution for a unified Communication Storage Interconnect, 
plus the physical rack 
and environmental structures required to deliver European Exascale Systems. (http://www.exanest.eu/)

EuroExa joins multiple European HPC
projects and partners with industrial SMEs (Small ad Medium Enterprises)
to co\text{-}design a groundbreaking supercomputing prototype. (http://www.euroexa.eu)
\subsection{Distributed Computing and Interoperable Data Access}
INAF-OATs participated, since the beginning, in Italian Grid and Cloud initiatives 
for the development of a 
multidisciplinary platform for distributed computing and data resources sharing.
Recently it participated in EU EGI.eu projects, coordinating the Astronomy and Astrophysics 
community, deploying a set of cloud provided IVOA compliant services with the aim to 
spread the IVOA standards based interoperability with CANFAR in Canada~\citep{cadcEgiInterop}.

INAF-OATs is involved in International Virtual Observatory Alliance (IVOA) leading the italian 
participation. Main activities in this area are the participation in the VO recommendation 
development; in the VO paradigm and tools dissemination to a wide audience (specialized and not); 
in the management and providing of web resources for the IVOA members collaboration.
\subsection{ESCAPE \& EOSC Integration}
ESCAPE (European Science Cluster of Astronomy \& Particle physics ESFRI research 
infrastructures) is a European H2020 project to integrate IVOA (International Virtual 
Observatory Alliance) compliant VO (Virtual Observatory) services within the EOSC (European 
Open Science Cloud) hybrid cloud scenario and to test containerization of VO aware applications~\citep{PosterMolinaroADASS2019}.
The ESCAPE project collects outcomes  of previous cluster projects ASTERICS (Astronomy ESFRI 
\& Research Infrastructure Cluster) and AENEAS (Advanced European Network of E-infrastructures 
for Astronomy with the Square Kilometer Array (SKA)). ASTERICS brought together researchers, 
scientists, specialists and engineers from astronomy, astrophysics and astro-particle physics in order
to develop instruments implementing common solutions to common challenges. AENEAS was to develop 
a science-driven, functional design for a distributed, federated European Science Data Centre (ESDC) 
to support the astronomical community once the Square Kilometre Array (SKA) becomes operational. 
Goal of ESCAPE is to integrate the results of these projects in a platform satisfying the 
requirements of SKA and the ESFRIs, ready to be part of the EOSC. The INAF-OATS computing 
facility will be used as an integration testbed in the scope of ESACAPE WP4 (Connecting 
ESFRI projects to EOSC through VO framework) and WP5 (ESFRI Science Analysis Platform) to 
integrate IVOA compliant VO standards and services within the EOSC hybrid cloud scenario 
and to test the containerization of VO aware applications. 
\section{Main scientific projects}
\subsection{LOFAR}
Since 2018 INAF is one of the partner to the International Low\text{-}Frequency Array Telescope
(LOFAR), a powerful radio telescope consisting of stations located in various countries in
Europe. It operates within the range 10-240 MHz, allowing detailed sensitive high\text{-}resolution
studies of the low\text{-}frequency radio sky and also providing an excellent short baseline coverage
to map diffuse extended emission. With its sensibility LOFAR is the most important SKA low
frequency precursor.

INAF\text{-}OATs implements and coordinates a distributed computing and data infrastructure in Italy 
to process LOFAR big data and to support scientists for the data reduction and analysis activities.
\subsection{EUCLID}
Euclid is an European Space Agency (ESA) space mission.  Scheduled for 2022, it will place a 
telescope in space with the aim of studying the properties of the dark universe. Euclid will 
collect many millions of images for at least 30 Pbytes of data, which will then have to be 
combined with other large data archives of images acquired with ground\text{-}based telescopes.
INAF\text{-}OATS is involved in the Euclid Consortium Science Working Groups and in the 
Euclid Consortium Science Ground Segment (ECSGS). It offers computing resources for the 
distributed computing Infrastructure of Euclid~\citep{euclid.spie.zac}.
\subsection{Numerical experiments}
The combination of a pre\text{-}exascale HPC infrastructure, joined by the development of 
novel paradigms for massively parallel computing applied to cosmological simulation codes, 
will enable scientists to carry out a multi\text{-}year simulation campaigns, whose final 
aim will be to provide a unifying interpretative framework for the cosmological experiments 
of the next two decades. A set of simulations interlaced in dynamic range (mass and force 
resolution) and cosmic time coverage would be designed to study cosmic evolution from the 
pre\text{-}ionization era, to the low\text{-}redshift universe. 2.0M core\_hours of 
INAF\text{-}OATS HPC cluster has been used for numerical experiments.
\section{Acknowledgments}
This work benefits support from the ESCAPE (grant n. 824064), EuroEXA (grant no. 754337) and ExaNeSt FET-HPC (grant no. 671553) projects funded by the European Union's Horizon 2020 research and innovation program.
\bibliography{P6-7}


\end{document}